\documentclass[universe,article,accept,pdftex,oneauthor]{Definitions/mdpi} 

\firstpage{1} 
\pubvolume{1}
\issuenum{1}
\articlenumber{0}
\pubyear{2023}
\copyrightyear{2023}
\datereceived{24 July 2023} 
\daterevised{31 August 2023} 
\dateaccepted{5 September 2023} 
\datepublished{ } 
\hreflink{https://doi.org/} 
\pdfoutput=1 

\usepackage{bm}

\def\p{\partial}

\def\ra{\rightarrow}

\def\Mfunction#1{\mathop{\rm #1}\nolimits}

\Title{Universality in the Exact Renormalization Group: 
Comparison to Perturbation Theory}

\TitleCitation{Universality in the Exact Renormalization Group: 
Comparison to Perturbation Theory}

\Author{Jos\'e 
 Gaite}

\AuthorNames{Jos\'e Gaite}

\AuthorCitation{{Gaite}
, J.}

\address[1]{%
Applied Physics Department
,
ETSIAE
,
Universidad Polit\'ecnica de Madrid, E-
28040 Madrid, Spain; jose.gaite@upm.es
}

\abstract{Various formulations of the exact renormalization group can be 
compared in the perturbative domain, in which we have reliable expressions for 
regularization-independent (universal) quantities. We consider 
the renormalization of the $\lambda\phi^4$ theory in three dimensions and make 
a comparison between the sharp-cutoff regularization method and other more recent methods. 
They all give good results, which only differ by small non-universal terms.}


\begin{document}


\section{Introduction}
\label{intro}

The exact renormalization group (ERG) 
\cite{Wil-Kog,Wegner-H,Polchi,Hasen2,Wett,Morris,Morris_1} 
is the non-perturbative formulation of the classical renormalization group, 
which was itself an improvement of perturbative quantum field theory. 
The ERG has been employed for the calculation of universal quantities for 
critical phenomena, in particular, critical exponents. Unfortunately, the ERG is afflicted by 
the problem of regularization scheme dependence of the results, which appears 
more acutely than in perturbation theory. However, considerable studies of 
regularization scheme dependence have been carried out \cite{Liao,Litim,Litim_1}, 
and Litim has proposed an optimized regulator
~\cite{Litim,Litim_1}, 
with which he 
obtains particularly accurate critical exponents~
\cite{Litim_CE}. 
At any rate, the comparisons made so far, for any given field theory, mostly involve 
the ERG fixed point and the perturbations about it (which provide the critical exponents). 
The ERG equations are actually applicable in a much larger parameter region. Thus, 
our intention is to explore other parts of the parameter space.

Undoubtedly, scale invariance is an important subject, and it is natural that many 
field theorists focus on the renormalization group fixed points. 
However, in many situations, one has to deal with field theories that are not scale invariant. 
Nevertheless, these theories need renormalization, 
which is normally implemented in perturbation theory \cite{QFT,Parisi,ZJ}. 
The classical renormalization group still is 
a convenient method of improving the results of perturbation theory, but the ERG opens a 
new avenue, given its non-perturbative nature. This idea was proposed 
years ago \cite{Shepard}, and it has recently been demonstrated numerically that 
the ERG equations can outperform perturbative renormalization within a range of coupling 
constant values \cite{I}. 

The regularization method employed in Ref.~\cite{I} is the simple sharp-cutoff method, which is said to be non-optimal \cite{Liao,Litim,Litim_1,Litim_CE}. 
However, the sharp-cutoff method is arguably useful within 
the local potential approximation \cite{Morris_4}. 
At any rate, reviews of the exact renormalization group written after the popularization 
of Litim's work \cite{KopietzBS,Delamotte,Rosten,Dupuis}
favor his method over the sharp-cutoff method. 
In a finite (truncated) coupling constant space, both methods lead to ordinary 
differential equations, whereas general smooth regulators lead to 
integro-differential equations \cite{Liao}. Ordinary differential equations are very 
suitable for numerical calculations and even lend themselves to some analytical 
investigations. 

One class of smooth cutoff functions consists of power-law functions with different exponents 
\cite{Liao,Litim,Litim_1}. In particular cases, these cutoff functions give rise to 
ordinary differential equations, as do the sharp-cutoff or the Litim cutoff functions 
\cite{Liao}. Among those cases, we shall pay attention to the 4th power law in 
three dimensions, which was originally studied by Morris \cite{Morris_1,Morris_3}. 
He concluded that it is more convenient than the sharp-cutoff function as a basis for 
the full derivative expansion of the ERG \cite{Morris_3}.

Usually, the ERG analysis of regulator performance is concerned with the ERG fixed points. 
Of course, the choice of regularization scheme affects the renormalization process 
for other  values of the coupling constants \cite{QFT,Parisi,ZJ}. Despite the fact that 
dimensional regularization has become standard in perturbative field theory, there is 
no fundamental objection to the use of other schemes in scalar field theories. Thus, 
one is prompted to ask for the effect that the various regulators already tested in the ERG
may have in the perturbative domain. 

One can consider several aspects of this question. 
The simplest field theory in three dimensions, the single-field scalar 
field theory, has only one non-trivial 
fixed point, namely, the Wilson--Fisher fixed point. This fixed point 
can be found employing the fixed-dimension perturbative $\lambda\phi^4$-theory  
(which actually provides very accurate values of the corresponding critical exponents)  
\cite{Parisi,ZJ}. 
One important fact to take into account is that this theory is super-renormalizable, 
that is to say, only a finite (and small) number of Feynman diagrams are 
\emph{superficially divergent} \cite{Parisi,ZJ}. 
When these diagrams are regulated, they produce regularization dependent terms, that is 
to say, non-universal contributions to the renormalized parameters. At any rate, 
the divergent diagrams only affect the renormalization of the mass $m$, whereas $\lambda$ 
is universal, because it does not involve divergences.

The relationship between the bare $\lambda_0$ and the renormalized $\lambda$ is well known 
in perturbation theory, to a high loop order, 
and it is employed to calculate the critical exponents of 
the Wilson--Fisher fixed point \cite{Parisi,ZJ}. This calculation indeed demands a high-order expansion and sophisticated resummation techniques. 
However, the perturbative series converges very well for small values of $\lambda/m$, so 
a few terms of it suffice to obtain very accurate results. 
Therefore, 
this fully perturbative region can be the adequate testing ground for a comparison with 
non-perturbative ERG results, in particular, regarding regulator optimality. 
Moreover, in addition to testing the relationship between $\lambda_0$ and $\lambda$, 
we can also test the relationship between $m_0$ and $m$, in spite of its not being universal.

Other aspects of the relation between truncations of the ERG equation and the 
perturbative renormalization group have been analysed before. Morris and Tighe 
\cite{Morris-Tighe} study the derivative expansion of the $\lambda\phi^4$-theory and 
compare the ERG beta function with the perturbative beta function to one and two-loop order. 
However, they focus on the massless case in four dimensions. 
In three dimensions, the perturbative series is hardly useful for the massless case, 
unless treated with sophisticated techniques, as noted above. 
Kopietz~\cite{Kopietz} employs Polchinski's ERG and initially keeps the dimension general, 
but he restricts the study, at some point, to $D \geq 4$. 
Kopietz's renormalization group equations are further complicated by his keeping the 
full momentum dependence. Here, we consider the local potential approximation of the ERG, 
which allows us to carry out simple calculations of the RG flow for the massive 
$\lambda\phi^4$-theory in three dimensions. 
Other articles that discuss the connection of the ERG with perturbation theory are 
Refs.~\cite{Litim-P,Codello-etal}, but they do not consider the 
$\lambda\phi^4$-theory in three dimensions.

Of course, the study of the connection between the exact and the perturbative 
renormalization groups is, generally speaking, as old as the theory of 
the renormalization group itself \cite{Wil-Kog}. It features in the early articles 
\cite{Hasen2,Morris} and early modern reviews of 
the \mbox{ERG~\cite{Aoki,Berges-Tet-Wet,Polonyi}}. However, these articles and reviews precede 
(or are simultaneous with) 
the studies of regularization-scheme optimization and are mainly concerned with 
the sharp-cutoff scheme, as the only one giving rise to tractable differential equations 
(it seems that Morris's scheme \cite{Morris_1}, namely, the 4th power law in 
three dimensions, was not sufficiently considered, perhaps because it is too specific).

A general study of the choice of regularization scheme in the ERG has been carried out 
by Baldazzi, Percacci, and Zambelli \cite{Balda}. In particular, they determine how 
the set of ERG beta functions depend on the choice of scheme and 
calculate and compare them for Litim's scheme and the $\overline{\mathrm{MS}}$ scheme 
in dimensional regularization. Some of the equations that we use here are, 
in fact, particular cases of their beta functions.
  
Our study consists of two parts. In the first one, spanning from Sections~\ref{sharp}--\ref{opt1}, we make a numerical comparison of three well-known ERG regularization methods with the universal perturbative renormalization formulas, including up to the 
three-loop order. The three regularization methods seem to work fine. 
In the second part, Section~\ref{reg&ren}, we make 
an analysis of the relationship between simple forms of the ERG differential equations for 
$m$ and $\lambda$ in the various schemes, on the one hand, and the one-loop gap and bubble equations, on the other. In Section~\ref{conclu}, we present some conclusions.
. 
 
\section{
Sharp-Cutoff Exact Renormalization Group and Perturbation Theory} 
\label{sharp}

Here we take the Wegner--Houghton sharp-cutoff ERG equations \cite{Wegner-H},
restricted to the single-field scalar 
field effective potential \cite{Hasen2,Morris_2}, in three dimensions, {namely}
, 
\begin{equation}\label{ERG}
\frac{dU_{\Lambda}(\phi)}{d\Lambda} = -\frac{A_3}{2} \,\Lambda^{2}\,
\ln\left[\Lambda^2 + U_{\Lambda}''(\phi)\right],
\end{equation}
where $A_3=(2\pi^2)^{-1}$ and $U_{\Lambda}(\phi)$ is the effective potential, such that 
$m^2=U_{\Lambda}''(0)$ and $\lambda=U_{\Lambda}''''(0)/4!$.
We compare the renormalization induced by Equation~(\ref{ERG}) 
to perturbation theory renormalization, for small 
values of the bare mass and coupling constant, where smallness is measured with 
respect to the only reference scale that we employ, which is the UV cutoff $\Lambda_0$.  
We allow negative values of $m^2$ and then smallness means 
$|m^2/\Lambda_0^2| \ll 1$.

As before \cite{I}, we 
employ the linearized Wegner--Houghton ERG for a rough approximation. 
Let us recall two simple conclusions from it: 
the coupling constant is not renormalized while $m^2$ grows as we lower the running cutoff 
$\Lambda$, namely,
\begin{equation}
m^2(\Lambda) = m_0^2 + \frac{6\lambda_0}{\pi^2}\,(\Lambda_0-\Lambda).
\label{mlin}
\end{equation}

Hence, a small positive renormalized mass (at $\Lambda=0$) requires $m_0^2<0$. Naturally, a better 
approximation, for example, a one-loop calculation or the 
non-perturbative approach of Ref.~\cite{I}, finds that $\lambda<\lambda_0$, and also finds a (negative) correction to Equation~(\ref{mlin}).

Let us make the mass and coupling constant non-dimensional by dividing each by 
the corresponding power of $\Lambda_0$. 
Notice that 
this non-dimensionalization is not of the usual type, which uses powers of 
the running cutoff $\Lambda$ \cite{Wil-Kog,Wegner-H,Hasen2,Morris_2}, 
but it is more convenient for us to compare with perturbative field theory results.
Nevertheless, our redefinition hides the fact that the mass renormalization 
is non-universal and, in particular,  
Equation~(\ref{mlin}) contains a term proportional to $\Lambda_0$ and divergent for $\Lambda_0\ra\infty$.
Let us leave the renormalization of mass for later 
and consider now the renormalization of the coupling constant 
in the perturbative domain, namely, for small absolute values of non-dimensional 
$m_0^2$ and $\lambda_0$. 

Thus, we first set $\lambda_0$ to some small number, say we set 
${6\lambda_0}/{\pi^2}=0.005$ ($\lambda_0=0.008225$).
Although Equation~(\ref{mlin}) gives only a rough approximation, we can use it 
to guide ourselves about the choice of $m_0^2$. 
Thus, let us take $m_0^2<0$ but $m_0^2>-{6\lambda_0}/{\pi^2}=-0.005$, because we want 
$m$ to be small, but not too small. 
We do not want to be close to masslessness (criticality) because  
it is not the truly perturbative domain. 
In addition, we must not take $m_0^2$ positive, especially positive and large, 
because then $\lambda$ is hardly renormalized (as occurs in the linearized ERG). 
We have tried $m_0^2=-0.0047 + 0.001 k,\;k=0,\ldots,5,$ and solved numerically the 
ERG equations, as we now explain.

The Wegner--Houghton ERG equations describe how  
the couplings in the effective potential flow with $\Lambda$. 
When truncated to a not-too-small number of coupling constants, 
the equations are known to be reasonably accurate, at least,  
for the analysis of critical behavior~\cite{Morris_2}. 
We employ them far from the Wilson--Fisher fixed point, namely, for 
non-vanishing but small initial values of $|m_0^2|$ and $\lambda_0$, and for 
initially vanishing values of the other couplings.

The numerical integration between $\Lambda=\Lambda_0$ and $\Lambda=0$ of the set of ordinary differential equations given by the 8th truncation of the Wegner--Houghton equation for the 
effective potential (up to $\phi^{16}$) yields the following results. 
For the renormalized mass and quartic coupling constant, we obtain ($k=0,\ldots,5$):
\begin{align}
m &= 0.009857,\, 0.03110,\, 0.04343, \,0.05312,\, 0.06138,\, 0.06869,
\label{mERG} \\
\frac{6\lambda}{\pi^2} &= 
 0.002575,\, 0.003774,\, 0.004051,\, 0.004196,\, 0.004291,\, 0.004359.
\label{lERG}
\end{align}

Hence,
\begin{equation}
\frac{\lambda}{m}=0.4298,\, 0.1996,\, 0.1534,\, 0.1299,\, 0.115,\, 0.1044.
\label{lm}
\end{equation}

These values are sufficiently small (except the first one) for us to keep 
a few terms of a series of powers of $\lambda/m$.  
Also note the relatively small variation of ${6\lambda}/{\pi^2}$ from its initial value 
${6\lambda_0}/{\pi^2}=0.005$ (except in the first case). 
Actually, Equation~(\ref{mlin}) roughly holds, as it~gives
$$m = 0.01732,\, 0.03606,\, 0.04796,\, 0.05745,\, 0.06557,\, 0.07280.
$$

As to the reliability of the 8th truncation of the ERG equations, 
we have checked that even truncations of somewhat smaller order yield essentially 
the same results.

For the comparison with perturbation theory, 
it is sufficient to keep up to $({\lambda}/{m})^2$ in the fixed-dimension perturbative series,  
that is to say, to keep up to the two-loop order in the renormalization of $\lambda$.  
In addition to this, we can also consider the sextic and octic coupling constants, 
which were calculated long ago in perturbation theory 
\cite{Soko-UO},  
and which we also obtain in our numerical integration of the ERG equations. 

To wit, the expressions that we employ are:
\begin{align}
\lambda_0 &= \lambda \left(1 + \frac{9 \lambda}{2\pi m}+ \frac{63\,\lambda^2}{4\pi^2 m^2}\right),
\label{l2loop} \\
g_6 &=  \frac{9 \lambda ^3 }{\pi m^{3}} \left( 1-\frac{3\,\lambda}{\pi m}
    + 1.389963 \,\frac{\lambda^2}{m^2}\right),
\label{g2loop} \\
g_8 &=  -\frac{81 \lambda ^4 }{2 \pi m^{3}} \left( 1-\frac{65\,\lambda}{6\pi\, m}
    + 7.775001\, \frac{\lambda^2}{m^2}\right),
\label{gg2loop}
\end{align}
where the non-dimensional sextic and octic coupling constants $g_6$ and $g_8$
refer to the terms next to $\lambda\phi^4$ in the expansion of the effective potential, 
namely, $g_6\phi^6+m^{-1}g_8\phi^8$
\cite{Soko-UO}. 

The values of $g_6$ and $g_8$ that we obtain with the ERG are: 
$$g_6 = 0.1389,\, 0.01728,\, 0.008275,\, 0.005176,\, 0.003658,\, 0.002774,$$
\unskip
$$-g_8 = 0.09582,\, 0.009248,\, 0.003793,\, 0.002127,\, 0.001380,\, 0.0009759.$$

The preceding perturbative formulas (\ref{l2loop})--(\ref{gg2loop}), 
in combination with Equation~(\ref{lm}), yield
\begin{align*}
\frac{6\lambda_0}{\pi^2} &= 
0.004919,\, 0.005093,\, 0.005093,\, 0.005090,\, 0.005088,\, 0.005087,\\
g_6 &= 0.1924,\, 0.01970, \,0.009166,\, 0.005652,\, 0.003959,\, 0.002983,\\
-g_8 &= 0.4195,\, 0.01272, \,0.004669,\, 0.002511,\, 0.001592,\, 0.001109.
\end{align*}

The comparison is successful, insofar as the ERG integration yields values of $\lambda/m$ and
values of $g_6$ and $g_8$ such that the substitution for $\lambda/m$
in the perturbative formulas 
approximately recovers the value of $\lambda_0$ and obtains values of $g_6$ and $g_8$ 
similar to the ones of the ERG integration. Naturally, the approximation is better the  
smaller $\lambda/m$ is, and the last values of $g_6$ and $g_8$ are off 
by about 10\%.

\section{Results for Litim's Optimized Exact Renormalization Group 
} 
\label{opt}

Here, we carry out the analogous calculations for Litim's optimally regulated ERG flow,  
employing his flow equation (Equation
~2.13 \cite{Litim_CE}). That flow equation applies to 
the $O(N)$ scalar field theory in $d$ dimensions, so we take the particular case 
$d=3$ and $N=1$, which, in terms of the dimensionful potential, {reads}
 \begin{equation}
\frac{dU_{\Lambda}(\phi)}{d\Lambda} = \frac{\Lambda^{4}}{6\pi^2\,[\Lambda^2 + U_{\Lambda}''(\phi)]}\,. 
\label{L-ERG}
\end{equation}

We again employ the 8th truncation of the flow equations. 
Litim studies the reliability of truncations (for fixed-point calculations) and finds that even lower order truncations are reliable 
(Section~3 \cite{Litim_CE}). We have also checked the reliability of the 8th truncation for 
our~calculations.

For the calculation of renormalized mass and coupling constants through 
Litim's flow equation, we need to set initial (``bare'') values. 
The linearization of Equation~(\ref{L-ERG}) yields
\begin{equation}
m^2(\Lambda) = m_0^2 + \frac{4\lambda_0}{\pi^2}\,(\Lambda_0-\Lambda).
\label{L-mlin}
\end{equation}

We can choose again (in dimensionless variables)
${6\lambda_0}/{\pi^2}=0.005$ ($\lambda_0=0.008225$), but now we require $m_0^2 > 
-{4\lambda_0}/{\pi^2} = -0.0033333.$
To have a set of initial values that can give results similar to the ones in the preceding 
section, we now choose \linebreak  $m_0^2= -0.00303333 + 0.001 k,\;k=0,\ldots,5.$
Nevertheless, we should not expect to recover the same values of renormalized mass, 
namely, the values in Equation~(\ref{mERG}). 
This does not matter, since our aim is to compare the results of the ERG integration with the results of the regularization-independent perturbative formulas (\ref{l2loop})--(\ref{gg2loop}), in which the bare mass $m_0$ does not feature. 
We only need $m$ and not $m_0$ in the perturbative formulas, 
and we only 
 need to assess to what extent the formulas are fulfilled.

The numerical integration is again straightforward and yields the following 
results:
\begin{align}
m &= 0.01083,\, 0.03146,\, 0.04370,\, 0.05335,\, 0.06158,\, 0.06889,
\label{mERG1} \\
\frac{6\lambda}{\pi^2} &= 
0.002620,\, 0.003770,\, 0.004053,\, 0.004202,\, 0.004300,\, 0.004370,
\label{lERG1}\\
g_6 &= 
0.1260,\, 0.01788,\, 0.008624,\, 0.005399,\, 0.003814,\, 0.002890,
\label{gERG1} \\
-g_8 &= 
0.08318,\, 0.009681,\, 0.004024,\, 0.002264,\, 0.001470,\, 0.001039.
\label{ggERG1}
\end{align}

Hence,
\begin{equation}
\frac{\lambda}{m}=
0.3979,\, 0.1972,\, 0.1526,\, 0.1296,\, 0.1148,\, 0.1043.
\label{lm1}
\end{equation}

In the present case, 
the relative variation of ${6\lambda}/{\pi^2}$ from its initial value 
${6\lambda_0}/{\pi^2}=0.005$ is not as small as before. 
Nevertheless, the values of $\lambda/m$ are small (smaller than before) and 
warrant the comparison with the low-order perturbative formulas.

Perturbative formulas (\ref{l2loop})--(\ref{gg2loop}), 
in combination with Equation~(\ref{lm1}), yield
\begin{align*}
\frac{6\lambda_0}{\pi^2} &= 
0.004775,\, 0.005069,\, 0.005089,\, 0.005095,\, 0.005097,\, 0.005099,\\
g_6 &= 
0.1517,\, 0.01901,\, 0.009019,\, 0.005606,\, 0.003943,\, 0.002979,\\
-g_8 &= 
0.2777,\, 0.01212,\, 0.004573,\, 0.002484,\, 0.001584,\, 0.001108.
\end{align*}

We find that the performance of this scheme is comparable to the one of the sharp-cutoff 
scheme.

\section{Results for Morris's Power-Law Cutoff Function
} 
\label{opt1}

Here, we carry out the calculations for Morris's power-law cutoff function,
employing his differential equation in $D=3$ (Equation~12 \cite{Morris_1}) in dimensionful form, 
{namely}
, 
\begin{equation}
\frac{dU_{\Lambda}(\phi)}{d\Lambda} = \frac{\Lambda^{3}}{2\pi\,[2\Lambda^2 + U_{\Lambda}''(\phi)]^{1/2}}\,. 
\label{M-ERG}
\end{equation}

We again truncate at $\phi^{16}$. 

The linearization of Equation~(\ref{M-ERG}) yields
\begin{equation}
m^2(\Lambda) = m_0^2 + \frac{3\lambda_0}{2^{1/2}\pi}\,(\Lambda_0-\Lambda).
\label{M-mlin}
\end{equation}

We choose again (in dimensionless variables) 
${6\lambda_0}/{\pi^2}=0.005$ ($\lambda_0=0.008225$), and now we require $m_0^2 > 
-{3\lambda_0}/(2^{1/2}\pi) = -0.0055536.$
Therefore, we set \linebreak  
$m_0^2=-0.0052536 + 0.001 k,\;k=0,\ldots,5.$ 
Our purpose is always to assess to what extent 
the perturbative formulas are fulfilled.

The numerical integration is again straightforward and yields the following 
results:
\begin{align}
m &= 
0.009914,\, 0.03109,\, 0.04341,\, 0.05310,\, 0.06135,\, 0.06866,
\label{mERG2} \\
\frac{6\lambda}{\pi^2} &= 
0.002531,\, 0.003748,\, 0.004030,\, 0.004178,\, 0.004274,\, 0.004343,\,
\label{lERG2}\\
g_6 &= 
0.1404,\, 0.01775,\, 0.008482,\, 0.005294,\, 0.003735,\, 0.002829,
\label{gERG2} \\
-g_8 &=  
0.09416,\, 0.009580,\, 0.003930,\, 0.002200,\, 0.001425,\, 0.001006.
\label{ggERG2}
\end{align}

Hence,
\begin{equation}
\frac{\lambda}{m}= 
0.4199,\, 0.1983,\, 0.1527,\, 0.1294,\, 0.1146,\, 0.1041.
\label{lm2}
\end{equation}

Perturbative formulas (\ref{l2loop})--(\ref{gg2loop}), 
in combination with Equation~(\ref{lm2}), yield
\begin{align*}
\frac{6\lambda_0}{\pi^2} &= 
0.004766,\, 0.005048,\, 0.005061,\, 0.005064,\, 0.005065,\, 0.005066,\\
g_6 &= 
0.1791,\, 0.01933,\, 0.009041,\, 0.005588,\, 0.003919,\, 0.002955,\\
-g_8 &= 
0.3700, 0.01239, 0.004587, 0.002474, 0.001572, 0.001096.
\end{align*}

The performance of this scheme is comparable to that of the others. 

\section{Regularization and Renormalization in the Exact Renormalization Group
} 
\label{reg&ren}

So far, we have tested the renormalization of coupling constants, which is \linebreak  
regularization-scheme independent. This is not the case for the relationship 
between $m_0$ and $m$, but this non-universal relationship is also worth considering. 
In this section, the role of $\Lambda_0$ is important, so 
we return to dimensional mass and coupling 
constants, that is to say, without dividing them by powers of $\Lambda_0$ 
(except when showing numerical results).

We have already derived first approximations to the mass renormalization in 
the three schemes considered, namely, 
Equations~(\ref{mlin}), (\ref{L-mlin}), and (\ref{M-mlin}).
However, we should not expect great accuracy from linearized ERG equations, which do not 
even renormalize the coupling constant. Fortunately, this method can be 
considerably improved by means of a simple non-perturbative formula, without 
considering coupling constant renormalization, namely, 
the ``gap equation'' \cite{Parisi}. Assuming a sharp cutoff $\Lambda_0$, the gap equation 
reads
\begin{equation}
m^2 = m_0^2 +  \int^{\Lambda_0}_0 
\frac{d^3k}{(2\pi)^3}\,\frac{12 \lambda_0}{k^2 + m^2}\,.
\label{cactusDS}
\end{equation}

This equation can actually be connected with the ERG \cite{Morris,Shepard,I}. 
In addition, it can be easily integrated to give (suppressing inverse powers of $\Lambda_0$): 
\begin{equation}
m^2= m_0^2 + \frac{6\Lambda_0}{\pi^2}\, \lambda_0 - \frac{3}{\pi}\, m\lambda_0\,.
\label{gapmr}
\end{equation}

This mass renormalization equation improves on
Equation~(\ref{mlin}) for $\Lambda=0$. For example, when solved for our values of $m_0$ and $\lambda_0$, 
it yields
$$
m = 0.01383,\, 0.03234,\, 0.04419,\, 0.05365,\, 0.06176,\, 0.06898.
$$

The agreement with the result of 
the numerical integration of the Wegner--Houghton ERG equation in Equation~(\ref{mERG})
is quite good. Further improvements can be achieved with the method 
employed in Ref.~\cite{I}.

We can also find an improved version of Equation~(\ref{L-mlin}) 
for Litim's regulator and reproduce the success above, in a certain sense. 
%
Let us first expound the 
connection between Equation~(\ref{cactusDS}) and the Wegner--Houghton ERG equation.
This equation admits an integral formulation, 
whose second derivative with respect to $\phi$ at $\phi=0$ {yields} 
 \cite{Shepard,I}:
\begin{equation}
m^2(\Lambda) = 
m^2(\Lambda_0) + 
12 \int^{\Lambda_0}_{\Lambda} \frac{d^3k}{(2\pi)^3}\,\frac{\lambda(k)}{k^2 + m^2(k)}\,.
\label{2DS}
\end{equation}

Taking $\Lambda=0$ and 
assuming that we can set 
$\lambda(k)=\lambda_0$ 
and $m^2(k)=m^2(0)=m^2$,
we obtain Equation~(\ref{cactusDS}).
Equation (\ref{2DS}) is equivalent to the differential equation
\begin{equation}
\frac{dm^2}{d\Lambda} = -\frac{6\,\lambda_0\,\Lambda^2}{\pi^2\,(\Lambda^2 + m^2)}\,.
\label{diffeq}
\end{equation}

This differential equation cannot be solved analytically (to our knowledge). However, 
when $|m_0^2|/\Lambda_0^2 \ll 1$, and as far as $|m^2|/\Lambda^2 \ll 1$, 
we can neglect the $m^2$ in the denominator, so we have 
a trivial differential equation, whose solution is Equation~(\ref{mlin}). 

Actually, Equation (\ref{diffeq}) derives from a truncation of the Wegner--Houghton 
ERG equation in which we assume $\lambda$
to be constant and, consistently, the higher-order coupling constants
to vanish.  
Thus, we are left with only 
the first equation of the hierarchy of ordinary differential equations. 
Of course, taking Litim's flow Equation (\ref{L-ERG}),
and under the same assumptions, we can also restrict ourselves to the 
first differential equation,  
which can be written as
\begin{equation}
\frac{dm^2}{d\Lambda} = -\frac{4\,\lambda_0\,\Lambda^4}{\pi^2\,(\Lambda^2 + m^2)^2}\,.
\label{diffeq1}
\end{equation}

(This equation is a particular case of the Litim scheme beta functions 
calculated by Baldazzi, Percacci, and Zambelli \cite{Balda}.) 
In analogy with Equation~(\ref{diffeq}), the solution of this equation, when 
$|m_0^2|/\Lambda_0^2 \ll 1$, is Equation~(\ref{L-mlin}). 
However, the integral equation that is equivalent to Equation~(\ref{diffeq1}) is {now}
\vspace{-6pt}
\begin{equation}
m^2(\Lambda) = m_0^2 + \int^{\Lambda_0}_{\Lambda}
\frac{d^3k}{(2\pi)^3}\,\frac{8\lambda(k)\,k^2}{[k^2 + m^2(k)]^2}\,.
\label{2DS1}
\end{equation}

Taking $\Lambda=0$, $m^2(k)=m^2(0)$, and $\lambda(k)=\lambda_0$, as we did in Equation~(\ref{2DS}), 
we now obtain,  
after integrating over $k$:
\begin{equation}
m^2 = m_0^2+\frac{2\lambda_0}{\pi^2} \left(3 \Lambda_0-\frac{\Lambda_0^3}{\Lambda_0^2+m^2}-3 m 
\arctan\frac{\Lambda_0}{m}\right)
\approx 
m_0^2+ \frac{4\lambda_0\Lambda_0}{\pi^2} -\frac{3\lambda_0\,m}{\pi}\,,
\label{gapmr1}
\end{equation}
where we have suppressed inverse powers of $\Lambda_0$ in the last expression.

Let us turn to Morris's power-law cutoff function. 
From differential Equation (\ref{M-ERG}),
we can derive, instead of (\ref{diffeq}) or (\ref{diffeq1}):
\begin{equation}
\frac{dm^2}{d\Lambda} = -\frac{6\,\lambda_0\,\Lambda^3}{\pi\,(2\Lambda^2 + m^2)^{3/2}}\,.
\label{diffeq2}
\end{equation}

Within the same approximation as above, we obtain, instead of 
(\ref{gapmr}) or (\ref{gapmr1}): 
\begin{equation}
m^2 \approx m_0^2+ \frac{6\lambda_0\Lambda_0}{2^{3/2}\pi} -\frac{3\lambda_0\,m}{\pi}\,.
\label{gapmr2}
\end{equation}

To interpret Equations~(\ref{gapmr}), (\ref{gapmr1}), and (\ref{gapmr2}), 
let us recall that the gap Equation (\ref{cactusDS}),
as the ``cactus approximation'' to the Dyson--Schwinger equation for the two-point function, 
is just an elaboration of the one-loop perturbation theory \cite{Parisi}.  
The one-loop mass renormalization is given by 
\begin{equation}
m^2 = m_0^2 +  \int 
\frac{d^3k}{(2\pi)^3}\,\frac{12 \lambda_0}{k^2 + m_0^2}\,.
\label{1loop}
\end{equation}

This integral is ultraviolet divergent, of course, and needs to be regularized. There 
are several methods of regularization in field theory, namely, modifications of the 
kinetic term in the action (or Hamiltonian), 
proper-time regularization, lattice regularization, \mbox{etc.~\cite{QFT,Parisi,ZJ}}. Usually, every method introduces a new parameter 
and gives a form of the integral that, in the divergent limit, 
can be split into a divergent term and 
a parameter-independent term. The latter term can also be calculated with methods that 
do not introduce a new parameter, such as subtraction methods or the method of 
differentiation.  
In fact, 
when we take the derivative with respect to $m_0^2$ of the integrand in Equation~(\ref{1loop}), 
we obtain a convergent integral, proportional to $(m_0^2)^{-1/2}$. 
The indefinite integral over $m_0^2$ recovers the divergent part as the 
arbitrary constant of integration and obtains the finite term 
$-{3m_0\lambda_0}/{\pi}$. This term is the one obtained with the economical methods of 
dimensional or analytic regularization and must be reproduced by every 
method of regularization. It is indeed common to 
Equations~(\ref{gapmr}), (\ref{gapmr1}), and (\ref{gapmr2}), although their respective 
divergent terms are different.


Finally, let us consider the one-loop perturbative renormalization of $\lambda$, first in 
connection with the second differential equation of the Wegner--Houghton equation hierarchy. 
This equation can be written as 
\vspace{-6pt}
\begin{equation}
\frac{d\lambda}{d\Lambda} = \frac{18\,\lambda^2\,\Lambda^2}{\pi^2\,(\Lambda^2 + m^2)^2}\,,
\label{WHdiffeq}
\end{equation}
where we have neglected the sextic coupling constant. 
To integrate this equation, 
let us assume that $m$ is constant (with $\Lambda$) and takes its renormalized value 
at $\Lambda=0$ (like we did to integrate Equations~(\ref{2DS}) or (\ref{2DS1}){):}
\begin{equation}
\int_{\lambda}^{\lambda_0}\frac{d\lambda}{\lambda^2} = 
\int_0^{\Lambda_0}\frac{18\,\Lambda^2\,d\Lambda}{\pi^2\,(\Lambda^2 + m^2)^2} = \frac{18}{\pi^2}
\int_0^{\infty}\frac{\Lambda^2\,d\Lambda}{(\Lambda^2 + m^2)^2} + 
\Mfunction{O}\!\left({\frac{1}{\Lambda_0}}\right)
= \frac{9}{2\pi m} + \Mfunction{O}\!\left({\frac{1}{\Lambda_0}}\right).
\end{equation}

This approximation, in the limit $\Lambda_0\ra\infty$, is equivalent to the so-called  
\emph{bubble\,approximation} 
 of the Dyson--Schwinger equation and 
obtains a simple expression of $\lambda_0$ as a function of $\lambda$ and $m$ (Equation~16 \cite{Shepard}),
namely,
\begin{equation}
\lambda_0 = \frac{\lambda}{1-9\lambda/(2\pi m)} 
= \lambda \left(1 + \frac{9 \lambda}{2\pi m}+ \frac{81\,\lambda^2}{4\pi^2 m^2} + \cdots\right).
\label{bubble}
\end{equation}

Of course, this function matches Equation~(\ref{l2loop}) to one-loop order. 

An expression equivalent to Equation~(\ref{bubble}) 
results from the classic renormalization-group-improved perturbation theory 
to one-loop order, with the beta-function \cite{Parisi,ZJ}
\begin{equation}
\left(m\frac{\p}{\p m}\,\frac{\lambda}{m}\right)_{\!\!\lambda_0} = 
\frac{\lambda}{m}\left(-1 + \frac{9 \lambda}{2\pi m}\right).
\label{b-fun}
\end{equation}

This beta-function does not refer to a flow with the cutoff $\Lambda$ but to 
the effect that a change of $m$ has on $\lambda$, once renormalization has been carried out, 
for a given value of $\lambda_0$.  
The integration of Equation~(\ref{b-fun}) between $m_1$ and $m_2$ yields:
\begin{equation}
\lambda_1 = \frac{\lambda_2}{1-(1/m_2-1/m_1)\,9\lambda_2/(2\pi)}\,. 
\label{b-fun-flow}
\end{equation}

We have seen that, for some $m_1$ large ($m_1 \gg \Lambda_0$), $\lambda_1(\Lambda)$ hardly changes 
with $\Lambda$; hence, $\lambda_1 \approx \lambda_0$ (the latter being its value at $\Lambda_0$). 
Therefore, for $m_1 \gg \Lambda_0 > m_2$, we neglect $1/m_1$ in 
Equation~(\ref{b-fun-flow}) and it becomes equivalent to Equation~(\ref{bubble}).

From Litim's equation hierarchy, in place of Equation~(\ref{WHdiffeq}), we have 
(see also Ref.~\cite{Balda}) 
\begin{equation}
\frac{d\lambda}{d\Lambda} = \frac{24\,\lambda^2\,\Lambda^4}{\pi^2\,(\Lambda^2 + m^2)^3}\,,
\end{equation}
whereas, from Morris's equation hierarchy, we have
\begin{equation}
\frac{d\lambda}{d\Lambda} = \frac{27\,\lambda^2\,\Lambda^3}{\pi\,(2\Lambda^2 + m^2)^{5/2}}\,.
\end{equation}

These two differential equations can be integrated with the same approximation 
made above, obtaining again Equation~(\ref{bubble}), in the limit $\Lambda_0\ra\infty$.


\section{Conclusions}
\label{conclu}

Our analysis of the Wegner--Houghton sharp-cutoff exact renormalization group 
equation demonstrates that it is a useful tool in the perturbative domain of $\lambda\phi^4$ 
theory in three dimensions. 
Unlike Morris \cite{Morris_4}, who cautions that, with the sharp-cutoff method in the local potential approximation,  
``truncations of the field dependence have limited accuracy and reliability'', we do find 
sufficient accuracy and reliability with a moderate truncation. 
This conclusion can be expected to hold for more general field theories. 

The good numerical concordance of 
the Wegner--Houghton sharp-cutoff ERG flow results with standard perturbative formulas 
holds in the other forms of the ERG flow that we study, namely, 
in Litim's or Morris's schemes. 
Moreover, a theoretical study of the effect of changes of the regularization 
scheme on universal magnitudes in standard renormalized perturbation theory 
leads us to unveil that Litim's or Morris's flow equations produce the same   
universal terms in the mass and coupling-constant renormalization formulas. 

Therefore, the difference between the various regularization methods lies in the 
non-universal terms, namely, the terms with explicit dependence on the parameter 
$\Lambda_0$. This is evident in regard to the mass renormalization, where the term 
proportional to $\Lambda_0$, divergent if $\Lambda_0\ra\infty$, is different in each case. 
Furthermore, the terms proportional to inverse powers of $\Lambda_0$, which are the only ones 
that appear in the coupling constant renormalization, are also non-universal. 
We have not calculated these terms, which produce corrections to universality 
of relative magnitude $m/\Lambda_0$. This amounts to corrections that are less than 10\% for our 
values of $m$, and this is indeed a bound to the order of magnitude of the deviations between the results of the three methods
that we find in the numerical integrations.

At any rate, the crucial test is the comparison of the results for universal quantities 
obtained by each method with the results of the perturbative formulas. 
In this comparison, we find again errors of less than 10\%, 
for sufficiently small values of $\lambda/m$. In addition, none of the three methods appears as 
definitely optimal. 

In conclusion, the choice of regularization method in this setting seems to be a matter 
of taste, to a large extent. The simple sharp-cutoff method is certainly a suitable choice. 
 


\vspace{6pt}

\funding{This research received no external funding.
}

\dataavailability{Not applicable.
}

\acknowledgments{I thank Sergey Apenko and the reviewer Gian Paolo Vacca for their comments.}

\conflictsofinterest{The author declares no conflict of interest.
}

\begin{adjustwidth}{-\extralength}{0cm}
\reftitle{References}

\PublishersNote{}
\end{adjustwidth}

\end{document}